# Checked-In Secret Detection: Strings Are All You Need

ZHENGDONG HUANG*, Southern University of Science and Technology, China
KEVIN LI*, McGill University, Canada
JINQIU YANG, Concordia University, Canada
YEPANG LIU, Southern University of Science and Technology, China
LILI WEI, McGill University, Canada

Hardcoded secrets in source code pose critical security vulnerabilities which can be easily exploited by malicious adversaries. Existing regex-based detection approaches suffer from fundamental limitations, as secrets often lack identifiable patterns, resulting in poor precision and recall. Recent studies have explored context-aware detection methods, as surrounding code can reveal the purpose of candidate strings. However, these methods confront three key challenges: (1) **obfuscation robustness** where models over-rely on easily obfuscated identifiers, (2) **cross-language generalization** difficulties due to uneven training data distribution, and (3) **lengthy and noisy context** that introduces excessive irrelevant tokens and slows inference.

We observe that **strings serve as a critical information source for code semantics**, offering superior contextual density, obfuscation robustness, and language independence. Based on this insight, we propose **StringGroup**, a new context extraction algorithm that mines strings surrounding potential secrets. By introducing a relatively simple modification to existing patterns that narrows the analysis specifically to string literals, the method achieves significant gains. With only **33.2%** of the original context, it preserves over **80%** of semantic information and significantly improves the signal-to-noise ratio for secret detection. We further design a context-aware secret detection tool, **Secretron**, based on StringGroup methods and Transformer model. Evaluation on the SecretBench dataset demonstrates high accuracy with **98.74% F1-score** and strong robustness under obfuscation and cross-language scenarios, outperforming state-of-the-art LLM-based baselines. We deploy our tool in real-world environments and successfully detect 48 previously unknown secret keys from 26 applications, demonstrating the practical effectiveness of our approach.

## 1 Introduction

Secret leakage has emerged as a pervasive threat across modern software ecosystems [72]. Developers often embed credentials, tokens, or cryptographic keys directly in code for convenience, enabling attackers to breach services with minimal effort. The GitHub Advisory Database [31]

*Zhengdong Huang and Kevin Li are co-first authors of this paper. This work was done when Zhengdong Huang was a visiting student at McGill University. Lili Wei is the corresponding author.

Authors' Contact Information: Zhengdong Huang, Southern University of Science and Technology, Shenzhen, China, 12212230@mail.sustech.edu.cn; Kevin Li, McGill University, Montreal, Canada, kevin.li3@mail.mcgill.ca; Jinqiu Yang, Concordia University, O-RISA Lab, Montreal, Canada, jinqiu.yang@concordia.ca; Yepang Liu, Southern University of Science and Technology, Shenzhen, China, liuyp1@sustech.edu.cn; Lili Wei, McGill University, Montreal, Canada, lili.wei@mcgill.ca.

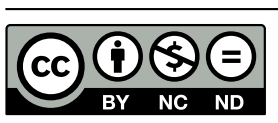

documents hundreds of potential security breaches caused by hardcoded secrets in open-source projects. In commercial environments, consequences can be devastating. For example, Uber's database secrets were once inadvertently uploaded to GitHub, exposing personal information of 50,000 drivers [16]. Such incidents highlight the urgent need for effective secret detection mechanisms.

To address this problem, numerous secret detection tools have emerged. The most prevalent approach employs regex-based pattern matching to identify potential secrets within code. Tools like TruffleHog [68], Detect-Secrets [76], and GitHub Secret Scanning [30] have gathered thousands of stars and been widely adopted in practice. However, such regex-based approaches face fundamental limitations because real-world secrets are typically highly random and often lack identifiable patterns. For instance, Twitter Client and Facebook secrets exhibit no consistent patterns [45], and many account passwords can be specified by developers with arbitrary strings. Regex-based methods require extensive pattern construction to handle this diversity, leading to over-generalization or over-specialization that causes low accuracy and requires substantial manual verification.

To address the limitations of regex-based approaches, prior works propose to leverage contextual information from the surrounding code. The key insight is that secrets of the same type are typically used in similar contexts even if they lack recognizable patterns. For instance, secrets of cloud services are often passed to cloud authentication APIs. Driven by the development of Natural Language Processing (NLP) in semantic understanding, various studies leverage code context to enhance detection accuracy [11, 25, 37, 48, 55, 58, 73]. These approaches employ machine learning models including Large Language Models (LLMs) to learn from code-snippet representations. Meanwhile, such representations face significant challenges for effective model learning:

- **Code obfuscation**: Previous research demonstrates that secret leakage occurs in obfuscated applications, where obfuscation eliminates semantic information such as function names and variable identifiers. Models learned from source code snippets over-rely on these easily compromised features, leading to poor robustness when detecting secrets in obfuscated code [20, 45].
- **Cross-language generalization**: Cross-language generalization is crucial as hardcoded secrets appear across diverse programming languages [8]. Despite similar underlying behavior, secrets appear within distinct representations and syntactic structures across languages. Models trained on specific languages rely heavily on language-specific features, compromising generalization to unseen languages, especially given imbalanced datasets favoring mainstream languages.
- **Lengthy and noisy context**: Code-snippet representations contain numerous generic tokens unrelated to code behavior, such as syntactic structures and boilerplate code. This creates lengthy representations with substantial noise while crucial secret-related information remains sparse. The low signal-to-noise ratio degrades computational efficiency and detection performance.

To address these challenges, our key observation is that **strings serve as a critical information source for reflecting code semantics**. In secret-related code such as cloud API calls or database connections, developers construct invocation payloads with multiple string literals that expose critical semantic information like service names, parameter keys, and authentication headers. String-based context extraction offers three advantages:

- **Obfuscation Resilience**: In obfuscated environments, identifiers such as function and variable names are typically transformed or randomized, causing significant information loss. However, string literals are commonly preserved and provide crucial contextual guidance even when other semantic indicators are compromised [20].
- **Cross-language Generalization**: String content transcends programming language boundaries. While syntax structures and keywords vary significantly between languages, string literals maintain consistent semantic meaning. This language-independent nature enables models trained on string-based features to generalize effectively across diverse languages.

- **Semantic Compactness**: Strings encapsulate rich semantic information within compact text spans. Unlike dispersed syntactic elements throughout code, strings naturally aggregate meaningful context directly related to application functionality. This concentrated semantic density enables efficient processing while preserving essential contextual information for detection.

Based on these insights, we introduce a relatively simple yet highly effective modification to existing context analysis techniques: narrowing the scope of context extraction specifically to string literals. We formalize this as **StringGroup**, a new context processing method that extracts semantic information to improve the accuracy of secret detection.

We systematically analyze the effectiveness of StringGroup on *SecretBench* [8], a large-scale real-world benchmark of secrets leaked on GitHub. Evaluations show that with only **33.2%** of the original code length, our approach retains **81.37%** semantic similarity while significantly increasing secret-related vocabulary frequency and outperforming baseline methods. StringGroup effectively preserves semantic information relevant to secret contexts while substantially reducing input length, thereby improving model performance.

Building upon StringGroup, we design **Secretron**, a well-designed, effective context-aware detection framework that fuses contextual information with intrinsic secret characteristics. For context understanding, we combine StringGroup with transformer architecture for robust semantic perception. To capture intrinsic secret patterns, we propose a parallel convolutional neural network that enhances string recognition. Our framework achieves **F1-Score 0.9874** on SecretBench, significantly outperforming existing methods. Notably, while conventional techniques and LLM-based baselines severely degrade under obfuscation and cross-language scenarios (most F1 < 0.77) , our approach maintains strong obfuscation robustness (**F1 0.9528**) and cross-language generalization (**F1 0.9290**). Experiments show that StringGroup effectively enhances LLMs' semantic perception capabilities for secret detection. Evaluation on Google Play [2] applications further identified 48 in-the-wild secret leakages from 26 applications, demonstrating the practical effectiveness of our approach in real-world scenarios. To summarize, we make the following contributions in this paper:

- **StringGroup Method**: We demonstrate that a relatively simple modification, which restricts the analysis entirely to string literals, can preserve core semantic information within concise tokens. We formalize this approach as a new context extraction method called **StringGroup**.
- **Context-aware Detection Framework**: Based on StringGroup, we propose **Secretron**, a well-designed context-aware secret detection framework integrating transformer and secret-specialized CNN encoders. On the *SecretBench* dataset, the model demonstrates high detection accuracy, strong robustness against obfuscation, and excellent cross-language generalization. Our framework depends on string extraction heuristics and source-level availability.
- **In-the-wild Evaluation**: We deploy our tool to scan closed-source applications from app stores, successfully identifying 48 in-the-wild secret leakages from 26 Android applications.

## 2 Motivating Example

To illustrate the limitations of existing approaches and motivate our solution, we present several concrete examples that highlight key challenges in secret detection. These examples demonstrate why context-aware methods require novel strategies to extract surrounding code, thereby enhancing robustness and generalization ability.

Listing 1 illustrates a hardcoded database access example from a Java application. Since database keys allow developers to specify arbitrary strings, regex-based methods fail to identify them accurately. Meanwhile, the code context provides relevant information about the key's behavior. As shown in lines 1-3 and lines 10-17, words such as "password" and "sql server" in the code clearly suggest that this string probably serves as a key for the SQL Server database connection. This

```
1  public static String dbName = "*********";
2  public static String administratorLogin = "
       *********";
3  public static String administratorPassword = "
       *********";
4  ...
5  .defineFirewallRule("allowAll")
6     .withIpAddressRange("0.0.0.1", "
          255.255.255.255").attach()
7  .defineElasticPool(epName).withStandardPool()
8  .attach().defineDatabase(dbName).
       withExistingElasticPool(epName)
9  .fromSample(SampleName.ADVENTURE_WORKS_LT).
       attach().create();
10 // Create a connection to the SQL Server.
11 Utils.print(sqlServer);
12 System.out.println("Creating␣a␣connection␣to␣the
       ␣SQL␣Server");
13 String connectionToSqlTestUrl = String.format("
       jdbc:sqlserver://%s:1433;database=%s;user=%
       s;password=%s;",
14     sqlServer.fullyQualifiedDomainName(),
15     dbName,
16     administratorLogin,
17     administratorPassword);
18 try (Connection conn = DriverManager.
       getConnection(url)) {
```

Listing (1) Example of hard-coded secrets in Java Project. Line 1-3 contains the hard-coded account and password for database authentication.

```
1  public static String a = "*********";
2  public static String b = "*********";
3  public static String c = "*********";
4  ...
5  a3 var1 = a2
6      .a("allowAll")
7          .a("0.0.0.1", "255.255.255.255")
8          .b()
9      .a(epName)
10         .a()
11         .b()
12     .a(a)
13         .a(epName)
14         .a(SampleName.ADVENTURE_WORKS_LT)
15         .b()
16     .a();
17 Utils.print((Object)var1);
18 System.out.println("Creating␣a␣connection␣to␣the
       ␣SQL␣Server");
19 String var2 = String.format("jdbc:sqlserver://%s
       :1433;database=%s;user=%s;password=%s;",
       var1.d(), a, b, c);
20 try (Connection var3 = DriverManager.
       getConnection(var2);){
```

Listing (2) The obfuscated representation of Code (1). It has comments removed and identifier names randomized. However, the string information in lines 18-19 is retained, and explicitly indicates that the code is used for database connection.

```
1  import (
2    "context"
3    "database/sql"
4    "fmt"
5    "log"
6    "time"
7    _ "github.com/go-sql-driver/mysql"
8  )
9  const (
10   dsn = "******:*********@tcp(*******:3306)/
         test_db?charset=utf8mb4&parseTime=True&
         loc=Local"
11 )
12 func InitDB() {
13     db, err := sql.Open("mysql", dsn)
14     if err != nil {
15         return nil, fmt.Errorf("open␣DB:␣%w",
               err)
16     }
```

Listing (3) Example of hard-coded secrets in Go Project. Compared to (1), they both contain similar authentication related strings like database connection URI (Line 12).

```
1  AIM(const std::string &ScreenName,
2  const std::string &Password,
3  const std::string &Toc_Server = "******",
4  const unsigned int &Toc_Port = 5190,
5  const std::string &Auth_Server = "******",
6  const unsigned int &Auth_Port = 5159,
7  const std::string &Agent = "CTocAim");
8  ~AIM(void);
9  AIMEvent do_event(void);
10 int send_im(const std::string &buddy, const std
       ::string &message);
11 int set_idle(const std::string &secs);
12 int set_idle(const unsigned int &secs);
13 int add_buddy(const std::vector<std::string> &
       buddies);
```

Listing (4) Example code of authentication handling in C Project. Type declarations and initializations (e.g., `const std::vector<std::string>`) occupy most of the code text. However, during code behavior understanding, this information is almost unrelated to the authentication logic.

provides crucial evidence to determine whether a string is a true secret. Despite the abundance of contextual information, effectively utilizing these clues remains challenging. We have identified three key challenges and demonstrate how our proposed **StringGroup**, a string-based context extraction method, can effectively address these challenges.

**Challenge 1 - Obfuscation Robustness**: A common approach to leverage contextual information is to input code snippets surrounding the candidate secrets, which has been widely adopted in previous studies [25, 48, 73]. However, this may cause the models to over-rely on comments and identifiers for decision making, which are not robust in many cases [20]. For example, Listing 2 shows the ProGuard-obfuscated version [35] of Listing 1. We observe that explicit comments indicating SQL server connection (line 10 in Listing 1) are removed during compilation, and meaningful variable names like the secret string identifier (line 1-3 in Listing 2) are replaced with meaningless symbols. In such cases, models cannot apply previously learned patterns based on identifiers or comments, which will lead to a significant degradation in accuracy and recall during prediction [45].

**Solutions**: String literals are relatively obfuscation-resilient features, thereby significantly enhancing model robustness under obfuscation scenarios. As shown in Listing 2, unlike function names and variable names that are typically obfuscated by most tools by default, string content remains robust against obfuscation scenarios. Furthermore, these strings provide important contextual clues such as request parameter names and functional descriptions in log messages. For example, lines 18-19 in Listing 2 explicitly describe the code's purpose for database connection, thereby facilitating code understanding.

**Challenge 2 - Cross-language Generalization**: Cross-language generalization is crucial in secret detection, as secrets can appear across diverse programming languages. However, projects in different languages have language-specific syntax, which can affect model decision-making and thereby reduce their generalization capabilities. Listing 3 illustrates a hard-coded database credential written in Go. While it performs the same SQL-server connection as the Java snippet in Listing 1, the connection string and API calls differ markedly. A detector overfitted to Java patterns therefore fails to identify this Go secret, revealing a critical gap in cross-language generalization.

**Solutions**: String content is relatively language independent, providing strong cross-language generalization capabilities. As demonstrated in Listing 1 and Listing 3, although these code segments are developed in Java and Go respectively, with different syntax rules and database connection representations, they both contain URLs required for constructing database connection requests (line 13 in Listing 1 and line 10 in Listing 3). Furthermore, their code contexts also include strings describing database names (line 12 in Listing 1 and line 13 in Listing 3), which strongly indicate that the code segments are used for database connections. This demonstrates the cross-language generalization capabilities inherent in string patterns.

**Challenge 3 - Lengthy and noisy context**: Furthermore, real-world source files can easily exceed the fixed token budgets of mainstream NLP encoders, yet a large fraction of those tokens is merely syntactic boilerplate with little semantic value. As shown in Listing 4, commonplace type declarations and container initializations (such as `const std::vector<std::string>`) occupy the bulk of the snippet, whereas authentication-related words appear only sparsely. These noisy tokens not only inflate the input sequence but also dilute the salient cues required for code understanding, thereby degrading the model's learning and inference efficiency. Therefore, a fine-grained context extraction strategy is required to filter out irrelevant components before model inference.

**Solutions**: String-based context provides comprehensive semantic information within shorter text spans. For Listing 4, the string "CTocAim" (line 7) clearly indicates that the code targets AOL Instant Messenger functionality. For Listing 1 as another example, we can infer the context is for SQL Server connection through the string "jdbc:sqlserver://%s:1433;database=%s;user=%s;password=%s;" without requiring complete code input. Furthermore, secrets frequently appear in authentication payloads or configuration files where they are surrounded by multiple strings. Therefore, we are able to extract sufficient string information.

## 3 Methodology

To develop a secret detection tool, we model secret detection as a discriminative problem. Specifically, we inspect each hard-coded string to determine whether it is a true secret. The overview of our proposed detection tool Secretron is shown in Fig. 1. We first heuristically extract candidate secret strings and their surroundings from the file, then employ the StringGroup method to preprocess the context. Subsequently, we utilized a dual-stream fusion classifier for end-to-end simultaneous secret detection and categorization. The classifier employs a transformer network and a parallel convolutional neural network to learn representations from both contextual and secret information, where these two streams are then fused for the final determination. We will introduce the details of StringGroup and the classifier in Section 3.1 and Section 3.2 respectively.

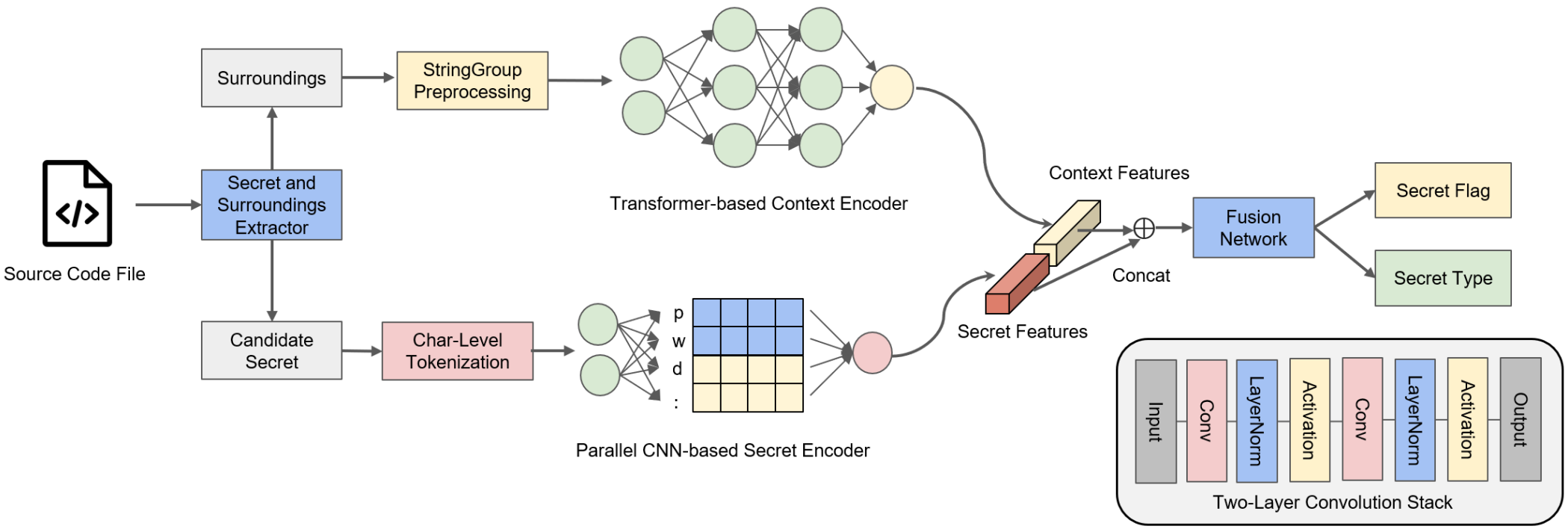

Fig. 1. Design of Secretron

## 3.1 StringGroup Method

**StringGroup** is a new lightweight context extraction method based on string literals. Leveraging the principle of program locality, it efficiently captures meaningful strings around candidate secrets while filtering out irrelevant noise. This ensures that the extracted features are both semantically rich and succinct for model training. The details of the StringGroup Algorithm are described in Section 3.1.1 and 3.1.2, and the method for selecting candidate secrets is described in Section 3.1.3.

*3.1.1 Overview of the StringGroup Algorithm.* The core algorithm iteratively mines strings in the vicinity of candidate keys, gradually expanding the collection window until the number of collected strings reach a threshold. Algorithm 1 presents the detailed procedure.

**Algorithm 1** The StringGroup Context Collection

**Require:** Source code $C$, candidate line $c$, max radius $R$, target capacity $S$
**Ensure:** Contextual string list $L$
1: $L \leftarrow [\,]$
2: **for** radius $r \leftarrow 0$ **to** $R$ **do**
3: $\quad Lines \leftarrow$ Valid lines at $(c - r)$ and $(c + r)$ {Radial expansion}
4: $\quad$ **for each** $line \in Lines$ **do**
5: $\quad\quad raw \leftarrow \text{ExtractStrings}(C[line])$ {Includes multiline compression}
6: $\quad\quad tokens \leftarrow \text{Preprocess}(raw)$ {See Sec. 3.1.2}
7: $\quad\quad L.\text{append}(tokens)$
8: $\quad\quad$ **if** $|L| \geq S$ **then**
9: $\quad\quad\quad$ **return** $L$
10: $\quad\quad$ **end if**
11: $\quad$ **end for**
12: **end for**
13: **return** $L$

The StringGroup algorithm starts from the center line containing the candidate ($r = 0$) and iteratively inspects adjacent lines. During extraction, any multiline strings are compressed into single lines to maintain parsing consistency. It then follows a three-stage pipeline for (1) collecting raw strings, (2) filtering out non-informative strings and (3) tokenization (Section 3.1.2), where the tokens will finally accumulate into list $L$. This outward expansion continues until either the maximum radius $R$ is reached or the collected tokens meet the target size $S$. We maintain the relative order of strings during the collection process to avoid disrupting semantic information.

*3.1.2 Preprocessing and Tokenization.* Following string collection, we apply specialized preprocessing to handle various string formats commonly found in source code:

*Filtering and Format Recognition.* The algorithm applies three key filtering and recognition steps:

(1) **High-entropy Filtering:** Strings with high entropy are filtered out and replaced with dedicated <RND> token. This is because the extracted context may contain a large number of highly random strings (such as UUIDs), which likely represent random character sequences that provide minimal semantic value. Since Shannon entropy serves as a good indicator of string randomness [61], we filter high entropy strings to reduce noise in context.
(2) **Length Filtering:** Words that are too short (e.g. only one char) are removed to eliminate noise. The purpose of this step is to reduce the number of low-information tokens [50] and improve the signal-to-noise ratio.
(3) **Special Format Recognition:** Code contexts often contain strings with distinctive patterns that correlate with secret occurrence, such as IP addresses and email accounts. To help models recognize these patterns more effectively, we employ regex matching to identify and extract special character formats including pure numbers, email addresses, IP addresses, and URLs. These are marked with dedicated tokens (<NUM>, <EMAIL>, <IP>, <URL>) appended to the end of the respective strings.

*Case Conversion and Tokenization.* We perform format conversion for `camelCase` and `PascalCase` strings by inserting spaces at word boundaries. Subsequently, we use LM's subword tokenizer [69] to provide finer-grained token segmentation. Finally, the tokenized outputs are used as the string contexts for the Context Encoder.

*3.1.3 Candidate Secret Determination.* To maximize recall, we adopt a permissive rule in our implementation: every string literal whose length is greater than six characters (without space) is treated as a candidate secret. This threshold ($> 6$) is based on the fact that most real-world secrets consist of continuous non-whitespace strings longer than six characters [34]. Alternative or more restrictive heuristics (entropy filters, keyword prefixes, etc.) can be integrated into our framework without modifying the subsequent StringGroup or model architecture.

### 3.2 Dual-Stream Fusion Classifier

We proposed a novel dual-stream fusion classifier that leverages both contextual information and intrinsic secret characteristics for accurate secret detection. Furthermore, with a dual-head architecture, the tool can not only determine whether a candidate string is a secret key but also identify its type. As illustrated in Figure 1, the architecture consists of two main components: a secret encoder that extracts features from candidate strings, and a context encoder that extracts features from the code surroundings preprocessed by StringGroup algorithm. Finally, the model includes fusion layers to combine the features from these two components

*3.2.1 Context Encoder.* The context encoder employs a Transformer-Encoder network [69] for secret context understanding. Following BERT [43], the <CLS> token is prepended to the tokenized string contexts to represent the entire sequence information for downstream classification tasks. The string context is then vectorized and processed through Transformer-Encoder Layers, producing transformed representations. The <CLS> token representation encapsulates comprehensive semantic information of the context features and serves as input for the subsequent model fusion stage.

*3.2.2 Secret Encoder.* The secret encoder processes each candidate string with a character-level convolutional neural network. We design a multi-scale, parallel convolutional block that extracts features at different contextual resolutions. Specifically, the block comprises several parallel convolutional blocks with distinct kernel sizes, and each block consists of convolutional layers stacked sequentially. After each convolutional layer, LayerNorm and activation modules are applied. These multi-parallel convolutional modules focus the model's attention on local segments of the string, helping extract various local features specific to secrets, such as the `sk-` or `aws-` prefixes. The extracted features are then concatenated into a single feature vector and subsequently transformed through a fully connected layer.

*3.2.3 Representation Fusion.* The fusion mechanism combines the `<CLS>` token representation from the context encoder with the secret representation from the secret encoder. These representations are concatenated into a unified feature vector and then processed through fully connected layers. The model finally split into two separate heads as outputs: one determines whether the string constitutes a secret key, and the other predicts its specific key type.

*3.2.4 Training Strategy.* In our implementation, we initialize the context encoder with Language Model's pre-trained weights [26], and the secret encoder is initialized with random parameters. During training, we simultaneously optimize the model's performance on both secret detection and secret categorization tasks. As it is a multi-objective optimization process, we adopt a weighted cross-entropy loss function:

$$\mathcal{L}_{total} = \underbrace{w_1\Big[-\sum_{i=1}^{N} y_i^{(1)} \log \hat{y}_i^{(1)} - (1-y_i^{(1)}) \log\big(1-\hat{y}_i^{(1)}\big)\Big]}_{\text{(1) Secret detection loss}} + \underbrace{w_2\Big[-\sum_{i=1}^{N}\sum_{c=1}^{C} y_{i,c}^{(2)} \log \hat{y}_{i,c}^{(2)}\Big]}_{\text{(2) Secret categorization loss}} \quad (1)$$

where $N$ is the number of training samples, $C$ is the number of secret types, $y_i^{(1)}$ and $\hat{y}_i^{(1)}$ represent the ground truth and predicted probability for binary secret detection of sample $i$, respectively, $y_{i,c}^{(2)}$ and $\hat{y}_{i,c}^{(2)}$ denote the ground truth and predicted probability for secret type $c$ of sample $i$, and $w_1$, $w_2$ are weighting coefficients that balance the contribution of each objective during training.

## 4 Evaluation

In this section, we comprehensively evaluate our tools for the following research questions (RQs):

- **RQ1 (Detection Accuracy):** How well does Secretron perform in detecting secrets?
- **RQ2 (Ablation Study):** How does each module contribute to the overall performance?
- **RQ3 (Robustness and Generalization):** To what extent does Secretron enhance obfuscation robustness and cross-language generalization ability?
- **RQ4 (In-the-wild Case Study):** How effectively does Secretron detect in-the-wild secrets?

### 4.1 Experiment Setup

*4.1.1 Dataset.* We conduct experiments using the SecretBench dataset [8], which contains 97,479 possible secret samples from 818 open-source repositories. To the best of our knowledge, this is the largest and most diverse secret dataset, covering multiple programming languages and secret categories. We randomly split the data at the repository level with a ratio of 7:1:2 into training, validation, and test sets. We further employed bootstrap resampling (iter=200) [22], which has been widely validated for reliable performance estimation in previous works [66].

*4.1.2 Evaluation Metrics.* We employ standard classification metrics: Accuracy, Weighted Precision (W-Prec), Recall (W-Recall) and F1-Score (W-F1) to assess model performance [24]. Additionally, we report positive class metrics (prefixed with "Pos-") to specifically evaluate the detection performance for actual secrets. We also conducted statistical tests to verify the significant differences between Secretron and other baseline models across various experimental settings, using the Wilcoxon Signed-Rank Test [74] and the Friedman test [29] with FDR Correction [10].

Table 1. Secret Detection Results

| Model | Acc | W-Prec | W-Recall | W-F1 | Pos-Prec | Pos-Recall | Pos-F1 |
|---|---|---|---|---|---|---|---|
| TruffleHog [68] | 0.5681 | 0.5803 | 0.5681 | 0.5511 | 0.6115 | 0.3735 | 0.4638 |
| Detect Secrets [76] | 0.5507 | 0.7107 | 0.5507 | 0.4454 | 0.8943 | 0.1150 | 0.2038 |
| LR | 0.6768 | 0.7860 | 0.6768 | 0.6427 | 0.9627 | 0.3679 | 0.5324 |
| SVM | 0.5694 | 0.7675 | 0.5694 | 0.4716 | 0.9976 | 0.1392 | 0.2443 |
| Naive Bayes | 0.5510 | 0.7240 | 0.5510 | 0.4437 | 0.5272 | **0.9904** | 0.6881 |
| KNN | 0.9011 | 0.9141 | 0.9011 | 0.9003 | 0.9875 | 0.8124 | 0.8914 |
| XGBoost | 0.9244 | 0.9334 | 0.9244 | 0.9240 | 0.9957 | 0.8525 | 0.9186 |
| PassFinder [25] | 0.7645 | 0.8396 | 0.7645 | 0.7507 | **0.9994** | 0.5293 | 0.6921 |
| CredSweeper [77] | 0.7769 | 0.7842 | 0.7769 | 0.7755 | 0.8298 | 0.6967 | 0.7575 |
| Rahman's Tool [55] | 0.9420 | 0.9421 | 0.9420 | 0.9420 | 0.9476 | 0.9357 | 0.9416 |
| Biringa's Tool [11] | 0.9073 | 0.9158 | 0.9073 | 0.9068 | 0.9750 | 0.8361 | 0.9002 |
| **Ours-Lite** | 0.9823 | 0.9823 | 0.9823 | 0.9823 | 0.9860 | 0.9785 | 0.9822 |
| **Ours-Acc** | **0.9874** | **0.9874** | **0.9874** | **0.9874** | 0.9924 | 0.9823 | **0.9873** |

*4.1.3 Implementation Details.* All experiments are performed on a single NVIDIA RTX 6000 GPU, using PyTorch 2.7.0 and CUDA Toolkit 11.4. All of the models are trained for 120 epochs with early-stop patience 10. We use the AdamW as optimizer with a learning rate of 0.001. Weight of different targets is set to $w_1 = w_2 = 0.5$. For StringGroup parameters, The radius $R$ of the context window surrounding the secret is set to 20, and max size $M$ of the extracted list is set to 500. We developed two tool versions with different pretrained context models: **Ours-Acc** uses DeepSeek-7B [4] for high accuracy, while **Ours-Lite** uses CodeBERT [26] with only 0.11B parameters for faster prediction while maintaining competitive accuracy.

## 4.2 RQ1: Detection Accuracy

In this section, we evaluate the overall performance of Secretron. Specifically, we focus on (1) determining whether candidate strings are true secrets, and (2) identifying their specific types.

*Baselines.* We compare it against existing widely used approaches across both regex-based and learning-based paradigms. For regex-based methods, we select TruffleHog [68] and Detect Secrets [76], both popular tools with over 4,000 GitHub stars. For learning-based methods, we select state-of-the-art context-aware approaches from previous research. These include PassFinder [25], a CNN-based secret detection tool, CredSweeper [77], a commercial tool that combines regex and BiLSTM for secret detection, and two LLM-based methods: Rahman's Tool (CodeLLaMA-7B) [55] and Biringa's Tool (GPT-2) [11], using their reported best-performing publicly available pretrained models. Most of existing tools, including above, focus on secret detection while neglecting fine-grained secret categorization. To ensure experimental validity, we strictly follow their original implementation and only compare with them in the detection task.

To better evaluate the effectiveness of our dual-stream classifier, we also propose several strong context-aware machine learning baselines. These baselines vectorize the candidate secret and its context and employ a *MultiOutputClassifier* for simultaneous secret detection and categorization. We use widely-adopted models including Logistic Regression (LR), Support Vector Machine (SVM), Naive Bayes, K-Nearest Neighbors (KNN), and XGBoost [12]. We compare the baselines with our proposed Ours-Lite (CodeBERT) and Ours-Acc (DeepSeek-7B). We adopt the classification metrics for evaluation as described in Section 4.1.

Table 2. Secret Category Classification Results (Weighted F1-Score)

| Model | Crypto Key | API Key | Auth Token | Generic Key | Database Key | Plain Password | Username | Other | W-F1 |
|---|---|---|---|---|---|---|---|---|---|
| LR | 0.6319 | 0.0331 | 0.0571 | 0.0000 | 0.0566 | 0.3607 | 0.0000 | 0.6453 | 0.4250 |
| SVM | 0.6107 | 0.0179 | 0.0074 | 0.0000 | 0.0262 | 0.3607 | 0.0000 | 0.6375 | 0.4065 |
| KNN | 0.8882 | 0.8970 | 0.9228 | 0.5985 | 0.9748 | 0.6966 | 0.9524 | 0.9110 | 0.9035 |
| Naive Bayes | 0.0394 | 0.0045 | 0.0249 | 0.0420 | 0.3384 | 0.2037 | 0.0042 | 0.1136 | 0.0838 |
| XGBoost | 0.9255 | 0.9392 | 0.9481 | 0.7482 | 0.9947 | 0.8000 | **1.0000** | 0.9388 | 0.9370 |
| **Ours-Lite** | 0.9755 | 0.9668 | **0.9723** | 0.8462 | **1.0000** | 0.8980 | **1.0000** | 0.9784 | 0.9741 |
| **Ours-Acc** | **0.9887** | **0.9723** | 0.9412 | **0.8861** | **1.0000** | **0.9000** | **1.0000** | **0.9934** | **0.9810** |

*Results.* **Secret Detection:** Table 1 presents the results for secret detection across our technique and all baseline methods. Our proposed approach demonstrates better performance across all evaluation metrics, achieving remarkable results with a weighted F1-score of **0.9823** for Ours-Lite and **0.9874** for Ours-Acc. Moreover, our method achieves high precision and recall rates (all >97.8%) for positive samples, indicating highly effective secret detection capabilities.

The results also reveal several key insights: First, regex-based methods like TruffleHog and Detect Secrets exhibit limited effectiveness, with F1-scores below 0.47, highlighting their inability to capture complex secret patterns. Second, while context-aware learning approaches show improved performance over regex-based methods, their performance remains inferior to ours (with $p < 1.2 \times 10^{-6}$ according to the Wilcoxon Signed-Rank Test), with the best baseline (ContextLLM) achieving an F1-score of 0.9420. This is because these methods fail to effectively capture the intrinsic features of both context and secrets, and do not attempt to fuse them. Specifically, existing LLM-based methods lack fine-grained feature modeling for secrets and rely on their coarse-grained sub-word tokenization, which could easily miss fine-grained patterns in secrets. In contrast, our dual-stream architecture employs the StringGroup algorithm to effectively enhance secret context comprehension, captures intrinsic secret patterns at adjustable scales, and efficiently fuses these information, leading to superior performance. Comparing two variations of our proposed method, Ours-Acc with a larger model achieves better accuracy than Ours-Lite, as more parameters could enhance contextual understanding. However, Ours-Lite has significantly lighter parameters (0.11B vs 7B), resulting in faster training and inference speeds. Therefore, users can select the appropriate version based on their specific scenarios, balancing accuracy and efficiency.

**Multi-class Secret Categorization:** Table 2 demonstrates our method's capability in fine-grained secret type classification. Our approach achieves outstanding performance with a weighted accuracy of 97.43% and an F1-score of 0.9741, significantly outperforming all baseline methods. With accurate secret category outputs, this approach facilitates further validation of key effectiveness and helps assess the impact scope of key leakage, thereby enabling better risk control of software.

**RQ1 Takeaway:** Secretron achieves superior secret detection performance with weighted F1-scores of 0.9823 (Ours-Lite) and 0.9874 (Ours-Acc), outperforming all baselines including state-of-the-art LLM-based methods. Users can choose between Ours-Lite (0.11B parameters, faster) and Ours-Acc (7B parameters, more accurate) based on their requirements.

### 4.3 RQ2: Ablation Study

*Overview.* To further investigate the contribution of each component in Secretron, we conduct fine-grained ablation experiments divided into three groups.

1. **Structural Ablation**: We systematically removed the context encoder and the secret encoder to examine how each channel's information contributes to the overall performance. We further ablated the design of each module including the convolutional kernel design in the Secret Encoder and the fusion method [38] between the two encoders.

Table 3. Secret Classification Ablation Result (Weight F1-Score)

| Metric | Origin | Structural Ablation | | | StringGroup Parameters | | | | | |
|---|---|---|---|---|---|---|---|---|---|---|
| | | W/O Ctx. | W/O Sec. | W/O StrG. | M=100 | M=200 | M=1000 | R=5 | R=10 | R=50 |
| W-F1 | 0.9823 | 0.8849 | 0.9213 | **0.9838** | 0.9576 | 0.9728 | 0.9810 | 0.9327 | 0.9753 | 0.9823 |

| Metric | Origin | Convolution Stack | | Fusion Methods | | | |
|---|---|---|---|---|---|---|---|
| | | Single Scale | Single Layer | And | Or | L.-Weight [38] | Cross -Attn. [69] |
| W-F1 | **0.9823** | 0.9533 | 0.9419 | 0.9196 | 0.8928 | 0.9557 | 0.9579 |

Table 4. Pretrained Model Ablation Result (Weighted F1-Score)

| Model | Origin | W/O StrG. | Model | Origin | W/O StrG. |
|---|---|---|---|---|---|
| DCodeBERT [60] | **0.9666** | 0.9533 | GPT-2 [54] | **0.9811** | 0.9748 |
| CodeT5 [71] | **0.9760** | 0.9641 | CodeLlama-7B [57] | **0.9868** | 0.9704 |
| Ours-Lite | 0.9823 | **0.9838** | Ours-Acc | **0.9874** | 0.9786 |

2. **Pretrained Model Ablation**: Since the context model uses pretrained language model, we further compared different pretrained models. We also ablated the use of the StringGroup method to verify its contribution to improving the model's performance.

3. **StringGroup Ablation**: We ablated key parameters in the StringGroup method, primarily including the context window radius $R$ and the maximum token $M$ to investigate their impact.

4. **String Filtering Ablation**: We further conduct a string filtering sensitivity analysis to examine the effectiveness of different minimum length thresholds and the alternative Shannon entropy-based filtering method [50].

*Analysis.* The ablation results are presented in Table 3 and 4, which reveals several key insights.

**Structural Ablation Analysis:** The structural ablation reveals the critical roles of both encoders. Removing the Context encoder (W/O Ctx.) causes the weighted F1-score to drop from 0.9823 to 0.8849, confirming that contextual information is important for secret detection. Similarly, removing the Secret encoder (W/O Sec.) reduces the F1-score from 0.9823 to 0.9213, demonstrating the necessities for modeling the intrinsic characteristics of secrets. While removing StringGroup yields a marginal gain ($\Delta < 0.0015$, Cohen's $d < 0.2$), this might be attributable to specific pre-trained weights, as StringGroup consistently benefits the majority of other pre-trained models (see below). For the convolution stack, performance degradation ($F1 < 0.954$) indicates that both core designs (multi-scale and dual-layer structure) are essential for capturing secret patterns of varying lengths. For the fusion method, all common alternative baselines from previous studies (commonly used AND/OR ensemble methods and advanced learnable-weight addition [38] and cross-attention fusion [69]) underperform our chosen feature fusion method. This validates the effectiveness of our key design choices.

**Pretrained Model Analysis:** For the pretrained model ablation, we selected different pretrained context models (DistilCodeBERT [60], CodeT5 [71], GPT-2 [54] and CodeLlama-7B [57]) following previous study. Results show that our model performs well across different pretrained models, with larger models yielding further accuracy improvements. Our method consistently outperforms prior LLM-based approaches under controlled comparison with identical pretrained models, demonstrating the effectiveness of our architectural design. Also, StringGroup improves prediction accuracy across most context models ($p < 0.01$), effectively enhancing input quality and performance.

**StringGroup Ablation Analysis:** Experiment shows that increasing the context window radius $R$ and the token count $M$ progressively improves the model's prediction accuracy. However, the precision gains gradually diminish as $R$ or $M$ increases. For the two experiments exceeding our preset parameters ($M = 1000$ and $R = 50$), the model shows no significant improvement, indicating that $M = 500$ and $R = 20$ provide an good balance between accuracy and computational efficiency.

Table 5. Comparison of Different Filtering Strategies

| Metric | Origin (len=6) | len=3 | len=8 | len=12 | Shannon-Entropy |
|---|---|---|---|---|---|
| Filtering Recall | 0.9989 | **1.0000** | 0.9913 | 0.9137 | 0.8141 |
| Remain Candidates | 231,187 | 1,457,295 | 218,337 | **190,816** | 224,402 |
| Model F1 | **0.9823** | 0.9151 | 0.9817 | 0.9148 | 0.8470 |

**String Filtering Ablation**: We evaluate four minimum length thresholds (len $\in$ 3, 6, 8, 12) following common constraints on secret strings [34], and include Shannon entropy [50] as an alternative. The thresholds are chosen around typical secret lengths of 6–8 [34], with 3 (0.5×6) and 12 (2×6) as lower and upper bounds. Table 5 reports filtering recall, retained candidate size, and final model F1 for each configuration. Our len= 6 achieves the best overall F1. Len= 3 retains all genuine secrets but produces an excessively large candidate set (6.3× ours), introducing noisy short strings that degrade model learning. Len= 8 performs comparably with only a marginal recall drop, while len= 12 discards many legitimate key strings, substantially suppressing final model recall despite high precision (0.9955). Shannon entropy retains a comparable number of candidates to len= 6 (within 3%), yet filtering recall drops significantly, as genuine and non-genuine secrets in dataset share similar entropy distributions and short invalid strings are not effectively removed. Overall, len= 6 best balances early-stage recall and noise reduction, yielding the best final performance.

*In-depth Analysis on StringGroup.* To further analyze how effectively StringGroup improves LMs' input quality, we employ statistical tools to examine the context features after processing. We use context word frequency visualization and statistical metrics to evaluate compression efficiency and semantic preservation.

**Metrics** To better understand context differences with and without applying StringGroup, we visualize word frequency distributions to compare vocabulary changes. The frequency is computed over the entire dataset using the following formula: $\text{Freq} = \frac{\text{number of occurrences}}{\text{number of total words}}$. To further evaluate how effectively StringGroup preserves information while achieving compression, we first use *compression rate* to measure the text length ratio before and after processing [62]. Second, we assess *text similarity* between original and processed texts, as similarity is widely used to assess information preservation [23]. We evaluate similarity from two perspectives using four mainstream metrics [28] (Table 6): semantic similarity measured by NLP indicators and word distribution similarity measured by statistical metrics. All metrics are normalized to [0,1], where compression rate closer to 1 indicates more compression, while other metrics closer to 1 indicate greater similarity.

Table 6. Overview of the Evaluation Metrics

| Formula | Name & Description |
|---|---|
| $\text{CR}(A,B) = 1 - \frac{\lvert B\rvert}{\lvert A\rvert}$ | **Compression Rate**: the reduction in text size. $A$ – original; $B$ – StringGroup; $\lvert\cdot\rvert$ – tokens sequence length. |
| $P = \frac{1}{\lvert C\rvert}\sum_{i\in C}\max_{j\in R} s(i,j), R = \frac{1}{\lvert R\rvert}\sum_{j\in R}\max_{i\in C} s(i,j)$ | **BERTScore** [78]. $P, R$ are precision and recall derived from the semantic-similarity matrix $s(i,j) = \cos(\mathbf{h}_i^c, \mathbf{h}_j^r)$ between candidate ($c$) and reference ($r$) embeddings obtained from BERT. The final score is the F1-score of $P$ and $R$. |
| $\text{SBERT}(u,v) = \cos(\mathbf{e}_u, \mathbf{e}_v)$ | **Sentence-BERT Cosine Similarity** [56]. $\mathbf{e}_u, \mathbf{e}_v$ – L2-normalised sentence embeddings of sentences $u$ and $v$ produced by Sentence-BERT; $\cos(\cdot,\cdot)$ – cosine similarity. |
| $\text{JSD}(P,Q) = \frac{1}{2}D_{KL}(P\Vert M) + \frac{1}{2}D_{KL}(Q\Vert M)$ | **Jensen–Shannon Divergence** [46]. $P, Q$ – probability distributions; $M = \frac{1}{2}(P+Q)$ |
| $H(P,Q) = \frac{1}{\sqrt{2}}\left\Vert\sqrt{P} - \sqrt{Q}\right\Vert_2$ | **Hellinger Distance** [39]. $P, Q$ – probability distributions |

Table 7. Top-20 Token Frequency before (Orig.) and after (Ours) StringGroup Preprocessing

| # | Ours ← | Freq(%) | Freq(%) | → Origin | # | Ours ← | Freq(%) | Freq(%) | → Origin |
|---|---|---|---|---|---|---|---|---|---|
| 1 | **amz** | 2.3130 | 2.1375 | type | 11 | **encryption** | 0.5857 | 0.9273 | name |
| 2 | **key** | 1.1802 | 2.1197 | locationname | 12 | side | 0.5535 | 0.8404 | **bucket** |
| 3 | type | 0.8448 | 1.9192 | header | 13 | align | 0.4786 | 0.8250 | nicc |
| 4 | **com** | 0.8405 | 1.8624 | location | 14 | **bigquery** | 0.4646 | 0.7857 | the |
| 5 | **bucket** | 0.7219 | 1.1988 | **amz** | 15 | size | 0.4394 | 0.7808 | return |
| 6 | parentname | 0.7140 | 1.1209 | this | 16 | this | 0.3900 | 0.7363 | float |
| 7 | the | 0.6445 | 0.9816 | **key** | 17 | **http** | 0.3843 | 0.7136 | dispname |
| 8 | NaN | 0.6067 | 0.9743 | parentname | 18 | members | 0.3575 | 0.6899 | structure |
| 9 | **server** | 0.5967 | 0.9675 | function | 19 | function | 0.3423 | 0.6881 | members |
| 10 | location | 0.5947 | 0.9664 | NaN | 20 | name | 0.3399 | 0.6480 | var |

Table 8. StringGroup Compressing Efficiency Measurement Results

| Metric | | StringGroup | Comment | Identifier | Syntax |
|---|---|---|---|---|---|
| **Length** | Compression Rate | 0.6680 | 0.8422 | 0.5556 | **0.9516** |
| **Semantic-Similarity** | BERT Score | 0.8821 | 0.5270 | **0.8949** | 0.7936 |
| | SBERT Cos-similarity | **0.7453** | 0.3418 | 0.5564 | 0.2006 |
| **Distribution-Similarity** | JS Similarity | **0.8015** | 0.6865 | 0.7162 | 0.1234 |
| | Hellinger Similarity | **0.5972** | 0.4860 | 0.4935 | 0.1047 |
| **Overall Metric Rank ($p$ = 0.0056)** | | **1.400** | 3.200 | 1.600 | 3.800 |

**Baselines** To quantitatively assess the efficiency, we conduct a comparative analysis against alternative code processing approaches by selecting three representative baselines [67]:

**Comment Extraction (Comment):** This baseline extracts only code comments, including single-line comments (e.g., `//`) and multi-line comments (e.g., `/* */`). Comments typically preserve semantic information about code's purpose and functionality through natural language descriptions.

**Identifier Extraction (Identifier):** This approach extracts only code identifiers like variable names and function names. Identifiers often carry semantic meaning through naming conventions and can provide insights into the code's domain and functionality.

**Syntax-Info Extraction (Syntax):** This baseline extracts syntax-related structures, including variable initialization, conditional statements, loop constructs, operators and function calls. These provided represents the logical structure and execution path of the program.

**Results** The *StringGroup* method effectively filters out grammar-related tokens and preserves critical security-relevant tokens, resulting in a higher signal-to-noise ratio.

Table 7 lists the top 20 most frequent tokens with (left) and without (right) StringGroup processing, where bold tokens are identified common secret-related words [8, 50]. After *StringGroup* processing, semantically meaningful and authentication-related tokens—such as `amz`, `key`, `com`, `bucket`, `server`, and `encryption`—move to higher positions. Furthermore, StringGroup effectively balances compression efficiency with information preservation, retaining **81.37%** semantic similarity and **69.96%** distributional similarity while compressing text to only **33.2%** of its original size. As shown in Table 8, StringGroup achieves the highest rank against other baselines, consistently outperforming them across most evaluation metrics. The Comment and Syntax baselines substantially compromise code information, which in turn results in significant semantic and distributional information loss. The Identifier baseline presents a strong comparison as identifiers typically contain functional indicators for variables and functions, yet it still underperforms StringGroup on the majority of metrics. These results demonstrate the effectiveness of the StringGroup method in code feature extraction.

Table 9. Model Efficiency Comparison

| Model | Params (B) | Inference Time (s/file) | TFLOPS | Training Time (min/epoch) |
|---|---|---|---|---|
| Ours-Acc | 6.91 | 2.3503 | 94.0878 | 105.83 |
| Ours-Lite | 0.12 | **0.1775** | **22.4753** | **10.97** |

*Inference Speed Analysis.* Since our framework provides two model variants trained on backbones of different scales, namely *Ours-Acc* (DeepSeek-7B) and *Ours-Lite* (CodeBERT-0.11B), we further analyze their efficiency under strictly identical environment configurations as described in Section 4.1.3. Table 9 reports the parameter size and average inference & training time for each variant. Overall, both variants achieve practical inference speeds suitable for real-world deployment. *Ours-Lite* is significantly more lightweight, with fewer parameters and lower Tera Floating Point Operations Per Second (TFLOPS), and runs approximately 10× faster in both training and inference compared to *Ours-Acc*. This distinction makes the two variants applicable to different deployment scenarios: *Ours-Lite* is well-suited for rapid CI/CD pipeline scanning where low latency is prioritized, while *Ours-Acc* is more appropriate for rigorous security audits that demand higher detection accuracy. Developers can therefore select the variant that best fits their operational requirements.

**RQ2 Takeaway:** Ablation studies confirm that the effectiveness of Secretron stems from our dual-stream architecture that effectively fuses well-captured intrinsic context and fine-grained secret features. Our design choices balance complexity and performance across all submodules, consistently outperforming prior LLM-based approaches under identical pretrained models. In-depth analysis reveals that StringGroup compresses text to 33.2% of its original size while retaining 81.37% semantic similarity, significantly enhancing context signal-to-noise ratio and improving prediction accuracy across context models.

## 4.4 RQ3: Robustness and Generalization

As suggested in previous sections, **obfuscation resistance** and **cross-language generalization** represent significant challenges for context-aware models. To investigate the robustness of our approach under these challenging scenarios, we design comprehensive experiments for evaluation.

### *4.4.1 Obfuscation Robustness Evaluation.*

*Experiment Setup.* For this evaluation, we select the five most prevalent languages from the SecretBench dataset: Python, JavaScript, Java, Go, and C++, and apply code obfuscation to their respective source files. We employ the most popular obfuscation tools for each language under their default settings. For languages whose obfuscators operate at the IR level (ProGuard for Java and OLLVM for C/C++), we additionally apply decompilers to restore a source-level representation before feeding the code into our context encoder (Table 10).

For baseline setup, since we have already shown in RQ1 that regex-based methods perform not ideally in secret detection, we primarily select competitive context-based methods as baselines. All tools are trained on the same unobfuscated training set and evaluated on the obfuscated test set. We ablated StringGroup algorithm as baselines (W/O StrG.) to further investigate its effectiveness.

Table 10. Obfuscation and Decompilation Toolchain Used in Our Study

| | Java | C/C++ | Go | JavaScript | Python |
|---|---|---|---|---|---|
| Obfuscator | ProGuard [35] | OLLVM [41] | GoObfuscate [17] | javascript-obfuscator [42] | python-minifier [27] |
| Decompiler | CFR [9] | Ghidra [52] | not-required | not-required | not-required |

Table 11. Classification Results under Obfuscation & Cross-Language Scenarios (Weighted F1-Score)

| | **Language** | **Baseline** | | | | **W/O StrG.** | | **Ours** | |
|---|---|---|---|---|---|---|---|---|---|
| | | PassF. | CredS. | R.Tool | B.Tool | Lite | Acc | **Lite** | **Acc** |
| **Obfuscation** | Python | 0.7203 | 0.4090 | 0.6957 | 0.6696 | 0.5811 | 0.6615 | 0.9150 | 0.9545 |
| | JavaScript | 0.5072 | 0.3754 | 0.8825 | 0.8315 | 0.7850 | 0.7509 | 0.9805 | 0.9852 |
| | Java | 0.7125 | 0.3507 | 0.6709 | 0.7542 | 0.5026 | 0.2740 | 0.9682 | 0.9672 |
| | Go | 0.8202 | 0.3658 | 0.7528 | 0.7737 | 0.7084 | 0.6619 | 0.9400 | 0.9250 |
| | C/C++ | 0.6814 | 0.3893 | 0.8071 | 0.6339 | 0.7352 | 0.7536 | 0.9605 | 0.9589 |
| | Origin Avg. | 0.8027 | 0.7591 | 0.9200 | 0.9204 | 0.9747 | 0.9815 | 0.9760 | 0.9841 |
| | Obfus. Avg. | 0.6883 | 0.3780 | 0.7618 | 0.7326 | 0.6624 | 0.6204 | 0.9528 | **0.9581** |
| **Cross-Lang** | PHP (4.99%) | 0.6156 | 0.7983 | 0.8286 | 0.7726 | 0.8521 | 0.8616 | 0.8914 | 0.9275 |
| | Ruby (4.77%) | 0.6576 | 0.5169 | 0.8197 | 0.7131 | 0.5778 | 0.9039 | 0.8934 | 0.9704 |
| | C# (4.39%) | 0.8521 | 0.6972 | 0.8128 | 0.8784 | 0.8535 | 0.9300 | 0.9339 | 0.9477 |
| | Scala (1.14%) | 0.9556 | 1.0000 | 0.9667 | 0.9217 | 0.9889 | 0.9670 | 0.9889 | 0.9783 |
| | Swift (0.40%) | 0.7793 | 0.7409 | 0.8745 | 0.7409 | 0.9062 | 0.6923 | 0.9373 | 0.9091 |
| | Cross-L. Avg. | 0.7720 | 0.7507 | 0.8605 | 0.8053 | 0.8357 | 0.8710 | 0.9290 | **0.9466** |

Table 12. Performance Comparison with Potential Alternative Methods.

| | Basic | Train from Obfuscated | Fine-tuning on Obfuscated | Train with StringGroup |
|---|---|---|---|---|
| Avg. F1-score | 0.6624 | 0.6563 | 0.6697 | **0.9528** |

We further evaluate common alternative methods that could potentially mitigate distribution shift and improve robustness (train/fine-tuning on obfuscated code).

*Results.* The obfuscation robustness results are presented in Table 11. The Origin.Avg and Obfus.Avg columns show the average F1 scores of five languages before and after obfuscation, respectively. Our method demonstrates exceptional robustness across all programming languages, maintaining near-original performance levels with an average weighted F1-score of 0.9528 (Ours-Lite) and 0.9581 (Ours-Acc) on obfuscated code, which corresponds to about 97% of the model's original performance. In contrast, all baseline models exhibit substantial performance degradation under obfuscation scenarios, with F1-scores dropping below 0.77 (Wilcoxon Signed-Rank $p < 1 \times 10^{-10}$). We also find that ablated baselines without StringGroup (W/O StrG.) suffer significant performance loss under obfuscation. This occurs because models trained on code snippets may rely heavily on comment and identifier information, which becomes irreversibly corrupted during obfuscation. We further evaluate two alternative distribution adaptation methods (train/fine-tune on obfuscated code) to examine whether basic models (W/O StrG.) can adapt to obfuscation scenarios. As shown in Table 12, models trained entirely on obfuscated code achieve worse performance, while fine-tuning on obfuscated code yields only marginal improvements. This is because the high noise and variability in obfuscated code pose challenges for model learning. In contrast, our StringGroup algorithm effectively extracts obfuscation-resistant features, enabling high performance even after obfuscation. This highlights the critical role of StringGroup for robustness under obfuscation.

#### *4.4.2 Cross-Language Generalization Evaluation.*

*Experiment Setup.* As hard-coded secrets can appear across diverse programming languages, we evaluate the cross-language generalization ability of our model by testing its effectiveness on unseen languages. For this experiment, we select ten languages with sufficient real secret samples for model training and evaluation. Specifically, we use the top five languages with the largest sample sizes— JavaScript (43.12% in the source code sample), Python (15.09%), Go (11.52%), Java (7.87%), and C/C++ (5.13%)—as the training set, and the five languages with relatively fewer samples—PHP (4.98%), Ruby (4.75%), C# (4.38%), Scala (1.13%), and Swift (0.40%)—as the test set for all models. Similar to previous sections, we also ablated StringGroup algorithm as baselines.

*Results.* Table 11 summarizes the classification performance under cross-language scenarios. Our model consistently outperforms all baselines across five less common languages with high F1 scores (Ours-Lite 0.9290 and Ours-Acc 0.9466). Despite never encountering these languages during training, the model effectively detects secrets within them. In contrast, most baseline methods suffer significant performance drops when generalizing to unseen languages (Wilcoxon Signed-Rank $p < 1 \times 10^{-10}$), which stems from context feature difference where many samples from unseen languages fall outside the decision boundaries learned from training data. The degradation of W/O StringGroup demonstrates that StringGroup extracts language-agnostic features, preventing overfitting to language-specific characteristics and enhancing cross-language generalization.

**RQ3 Takeaway:** Our method demonstrates exceptional robustness under challenging scenarios while all baselines suffer significant degradation. For obfuscation resistance, Secretron maintains 97% of its original performance on obfuscated code (F1-scores Lite: 0.9528 and Acc: 0.9581). For cross-language generalization, our model achieves high F1-scores (Lite: 0.9290 and Acc: 0.9466) on unseen languages. Ablation studies confirm that StringGroup plays a critical role in both scenarios by extracting obfuscation-resistant and language-agnostic features, thereby ensuring robustness in obfuscation and cross-language generalization for secret detection.

### 4.5 RQ4: In-the-wild Case Study

*Experiment Setup.* To better understand the effectiveness of our method in real-world scenarios, we conducted a large-scale experiment to evaluate its practical applicability in detecting secrets within Android applications.

**Baselines:** Following previous researches [45], we adapted our Secretron-Lite to Android APK scanning. We selected LeakScope [80] and Three-layer Filter [50] as comparing baselines. These were strong tools evaluated in previous research and were specifically adapted [45] to the Android secret detection scenarios. We did not use the other existing techniques as baselines in this RQ since they exhibited significant performance degradation in obfuscated scenarios as demonstrated in RQ3, and existing LLM-based methods are prohibitively slow for large-scale evaluation [13].

**Dataset:** To evaluate our tool in the latest Android ecosystem, we scanned 900 Android APKs collected from the Google Play Store [2] in July 2025. To mitigate popularity bias, we built the dataset by randomly sampling words from the DWYL English dictionary [21] and collecting the ten most relevant results per query, following prior large-scale app studies [19]. Given that prior work reports a secret prevalence of 5%–20% in Android apps [51], a sample of $n = 900$ yields a margin of error of ±2.6% at 95% confidence via Cochran's formula [15], sufficient for statistically reliable

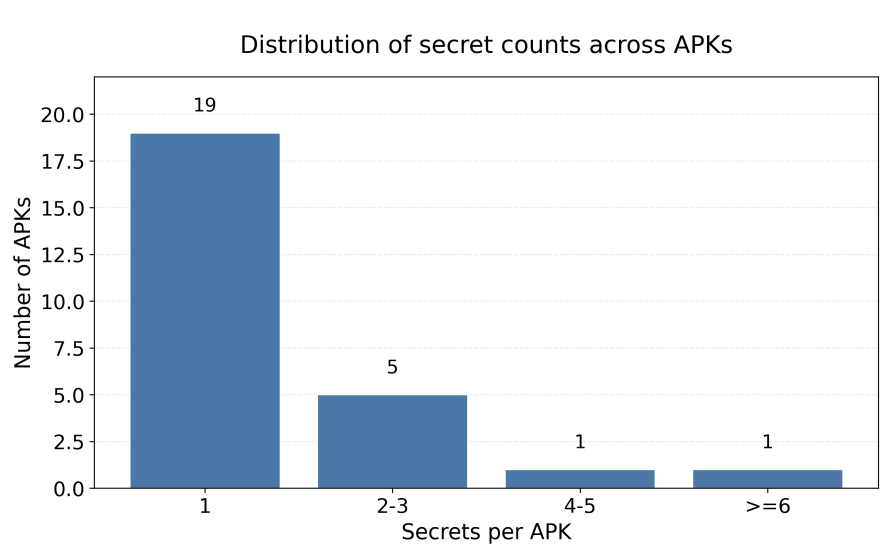

(a) Distribution of secret counts across APKs.

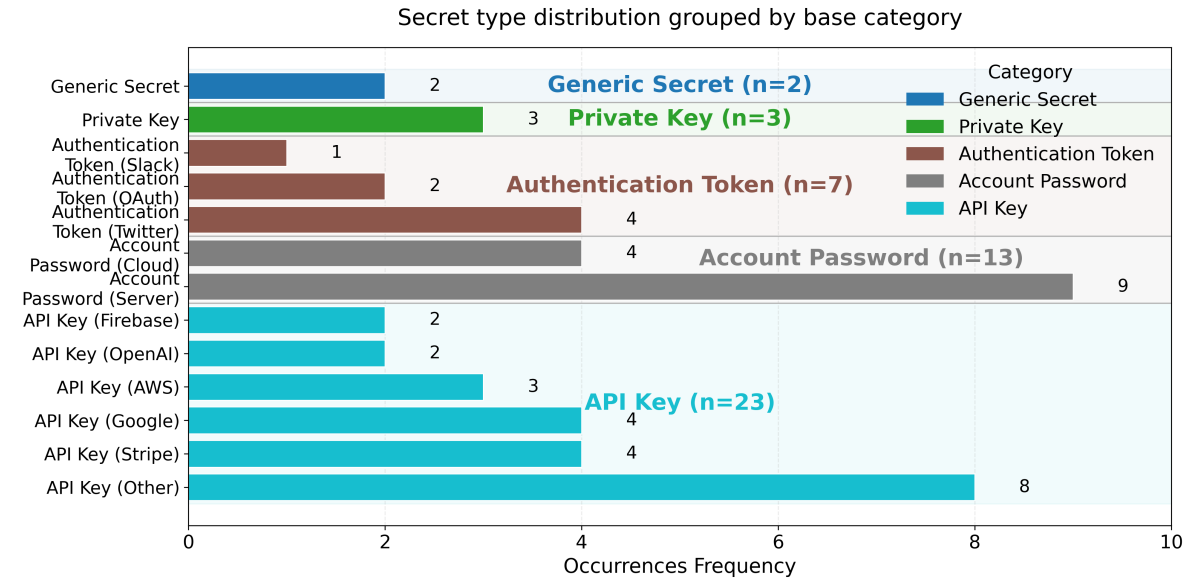

(b) Grouped bar chart of secret types.

Fig. 2. Overall performance and distribution analysis of the proposed method.

Table 13. Detection Performance for Android Applications

| Model | #Warnings | TP | FP | Precision | Precision 95% CI Range |
|---|---|---|---|---|---|
| LeakScope | 1 | 1 | 0 | **100.0%** | [0.0250, 1.0000] |
| Three-Layer Filter | 100,322 | 1 | >100,000 | <1% | [0.0000, 0.0001] |
| **Secretron (Ours)** | 54 | **48** | 6 | 88.9% | [0.7737, 0.9581] |

conclusions. Applications in the Google Play Store are primarily commercial, with approximately 43% being obfuscated [20]. We then scanned these APKs using Jadx [64] as the decompiler.

**Evaluation Metrics:** Two software security engineers with over three years of developing experience independently examined each detected secret's context and usage patterns, achieving high agreement (Gwet's AC1 [36] score 0.874). They then discussed ambiguous cases to reach consensus. Strict criterion were applied when labeling: secrets are genuine only when they match valid patterns [50] and are used in security-critical operations such as API authentication or cryptographic operations. We computed metric bounds using the Wilson confidence interval at 95% confidence level [14]. It is hard for recall evaluation due to the lack of ground truth. Thus, we primarily evaluate the number of true positives discovered by different tools and their precision.

*Results.* Our experimental results in Table 13 demonstrate promising detection capabilities in real-world applications. Specifically, 48 out of 54 secrets (88.9%) from 26 APK files were verified as genuine leakage, while the remaining were highly suspicious but lacked identifiable usage. In contrast, LeakScope detected only one genuine secret also found by our tool. Three-Layer Filter produced numerous detections which were impossible to evaluate all. Thus, we deduplicated the results and match the exact secret values, finding that only 1 of the 48 genuine secrets identified by *Secretron* appeared in the Three-Layer Filter's results, demonstrating that our tool detects more valid secrets than baselines. Furthermore, we randomly sampled 300 instances for manual inspection—a sample size that achieves 95% confidence with a margin of error below ±5% per Cochran's formula [47]—and found none to be genuine secrets, indicating its precision below 1%.

**Distribution of Secrets:** As shown in Fig. 2, among the 26 applications with genuine leaks, 19 (73%) contained exactly one secret, 5 (19%) contained 2–3, and 2 contained 4 or more, with one outlier holding 15 secrets due to direct client-side aggregation of multiple third-party services. Following the categorization of Basak et al. [8], the 48 secrets span five categories: API Keys were the most prevalent (n=23, 47.9%), alongside passwords and token for authentication. Cloud credentials spanned multiple providers (Google, Amazon, Stripe, Firebase, and OpenAI), demonstrating broad coverage of modern secret types including emerging LLM service credentials.

**In-depth Case Study:** Our tool remains effectiveness in heavily obfuscated scenarios. The example in Listing 5 shows a popular application (5M downloads) written in Kotlin with heavy obfuscation

```
1  @Override // kotlin.lqa
2  public int B1() {
3      return (int) this.config.f("feature.
           picture_feed.chunk_size", 30L, 10L,
           100L);
4  }
5  @Override // kotlin.lqa
6  public String C() {
7      return this.config.j("feature.camera.
           recording.quality", "HD");
8  }
9  @Override // kotlin.lqa
10 public String C0() {
11     return this.config.j("fake.buttons.
           slack_webhook.token", "xoxb
           -*************************");
12 }
13 @Override // kotlin.lqa
14 public long C1() {
15     return this.config.e("location.
           maximalAccuracy", 500L);
16 }
```

Listing (5) Secret detected in decompiled code from an obfuscated Android Kotlin project (Line 11).

```
1  public final class Urls {
2      public static final String BASE_URL = "https
           ://shopifymobileapp.cedcommerce.com/";
3      public static final String BIRTHREWARDS = "
           https://loyalty.yotpo.com/api/v2/
           customer_birthdays";
4      public static final String gpturl = "https
           ://api.openai.com/v1/completions";
5      public static final String MulipassSecret =
           "*********************";
6      private static String authtoken = "Bearer␣sk
           -**********************";
7      private static String content_gpt = "
           application/json";
8      private static String model = "text-davinci
           -003";
9      private static double temperature = 0.7d;
10     private static int max_tokens = 64;
11     ...
```

Listing (6) A configuration class extracted from the Android project contains 35 authentication-related hardcoded items (APIs, accounts, secrets), including 15 secrets spanning 13 cloud service providers.

that obscures identifier semantics. However, string contexts still preserve critical context: strings like “feature.camera.recording.quality” indicate a configuration file while “slack_webhook.token” directly exposes the adjacent Slack token. The StringGroup method effectively extracts these string features to detect it, maintaining detection capability in obfuscated codes.

Our tool can further detect secret types beyond the training set. Example in Listing 7 exposes administrator credentials for KangarooRewards, a less-popular third-party service. This case challenges existing methods because the password lacks typical patterns and KangarooRewards is absent from existing secret datasets. Nevertheless, our tool successfully identified this secret through strongly correlated contextual information such as associated email addresses. This demonstrates robust context-awareness and the ability to generalize beyond training data.

We also observed that secrets frequently appear in configuration classes with rich semantics. As shown in Listing 6, a single configuration class contained 35 authentication-related hardcoded items, including 15 secrets spanning 13 cloud service providers. This concentration makes secrets particularly susceptible to detection through our contextual analysis approach.

```
1 private static String KangarooRewards_ADMIN_ID = "s**********@******.com";
2 private static String KangarooRewards_ADMIN_PASSWORD = "1***************a";
3 private static String KangarooRewards_CLIENT_ID = "1*****1";
4 private static String KangarooRewards_CLIENT_SECRETE = "4******************b";
```

Listing 7. Hardcoded password for logging into the backend account detected in the Android project.

> **RQ4 Takeaway:** Our evaluation on 900 real-world Android applications demonstrates high precision of 88.9% with 48 verified genuine secrets detected from 26 APKs. Case studies reveal that Secretron successfully detects secrets in heavily obfuscated applications and out-of-training-set secret types, confirming its practical applicability and real-world robustness.

## 5 Discussion

### 5.1 Threats to Validity

**Dataset Annotation Quality.** Existing secret-related datasets are collected with manual annotations, which may contain labeling errors impacting evaluation. To minimize this threat, we utilized the SecretBench [8], the largest available dataset with over 90,000 samples. We further deployed our tool in in-the-wild scenarios and successfully detected actual secrets.

**Recall in Real-World Study.** A primary limitation of our real-world study is the inability to measure absolute recall, as complete ground truth for genuine secrets in in-the-wild Android APKs is inherently hard to obtain. This limitation affects the strength of our empirical claims by restricting our conclusions to relative performance rather than absolute detection recall. Therefore, we evaluate the intersection of true positives detected by Secretron and the baselines, which demonstrates that Secretron identifies a larger volume of confirmed secrets than baselines, achieving higher relative recall while sustaining well precision.

**Ethical Considerations.** Secret detection research inherently carries ethical risks requiring careful consideration. To mitigate potential harm, we refrained from verifying whether detected secrets are functional, as validation could damage developers. Instead, we manually examined all results to determine legitimacy. We are notifying developers about identified leakage through responsible disclosure, ensuring vulnerabilities are addressed while minimizing exploitation risks.

### 5.2 Limitations of Our Approach

**String Encryption.** String encryption is an obfuscation techniques that uses cryptographic methods to encrypt string literals, making them appear encrypted during reverse engineering. However, several factors significant mitigate this threat to our tool:

- **Limited adoption**: String encryption is rarely used in mainstream applications. Dong et al. [20] found that while 40-70% of applications use identifier renaming in app markets, only less than 2% employ string encryption. This is because string encryption significantly increases runtime overhead, leading tools like ProGuard to omit or disable it by default.
- **Incomplete coverage**: Not all strings can be encrypted. For instance, external resource loading requires plaintext strings to maintain functionality, limiting encryption deployment [32, 53].
- **Reversible implementations**: Many string encryption tools use statically hardcoded decryption functions [5, 40, 49, 65]. Since the decryption algorithm is embedded, static analysis or dynamic capture methods can recover decrypted strings. Previous studies have demonstrated feasibility for mainstream encryption tools [18, 33].

Furthermore, we acknowledge that advanced obfuscation techniques such as VM-based obfuscation and OLLVM extensions [70] provide stronger protection for string literals and reduce the effectiveness of string-based detection. However, such advanced obfuscations are adopted in few applications due to performance overhead [20]. To address well-protected binaries, we leave the integration of advanced deobfuscation tools [18, 32] as future work.

**Regex-based String Extraction.** Our approach still relies on regular expressions for syntactic string extraction. If a future language adopts a novel string representation paradigm, a new extraction rule must be added. However, this effort is inherently minimal: unlike secret detection, where secrets lack fixed patterns and tools such as Trufflehog [68] require 750+ handcrafted expressions yet still achieve limited recall [6], string literals share consistent syntactic patterns across languages, and 9 universal rules suffice to cover 15+ languages in our benchmark. Future work could adopt language-agnostic parsers such as Tree-sitter to eliminate this residual overhead entirely.

**Line-based Context Extraction:** We opted for line-based extraction to maintain a lightweight and universally applicable framework. Although its practical robustness is validated on SecretBench, relying on physical lines makes the method potentially vulnerable to complex representations. Future work could explore structural representations (e.g. ASTs) to further enhance the robustness.

**Decompilation Dependency.** Our approach inherently relies on source-level availability. However, analyzing closed-source software such as Android applications requires decompilation, and its effectiveness is therefore bounded by decompiler quality. Future work should explore byte-level semantic extraction for better cross-program extension.

## 6 Related Work

**Secret Leakage:** Extensive research has investigated secret leakage across various platforms and evaluated detection tool performance in different contexts. Basak et al. [6] examined regex-based tools on GitHub open-source code. Wei et al. [72] and Li et al. [44, 45] investigated Android secret leakage and evaluated detection tools on mobile platforms. Yadmani et al. [75] focused on secret leaks in cloud APIs. These studies reveal that secret leakage is pervasive in both open-source and closed-source applications. Moreover, leaked secrets are proved to be numerous, easily exploitable, and can cause devastating consequences. This poses severe threats to software security, resulting in critical risks including large-scale data breaches and widespread service disruptions [44, 72].

**Secret Detection:** Due to widespread secret exposure risks, secret detection has become a critical research area. Popular open-source tools like TruffleHog [68] and Detect-Secrets [76] have garnered thousands of stars on GitHub and are widely adopted. However, Basak et al. [6] demonstrate that such regex-based approaches suffer from low precision and recall. Research has evolved from basic pattern matching to context-aware approaches. Early work focused on hand-crafted features for context understanding. Saha et al. [58, 59] designed regular expressions with contextual features to reduce false positives. Lounici et al. [48] and Wen et al. [73] further enhanced feature engineering. Recent methods leverage deep learning to understand code context semantics [7, 25, 37, 63, 79], with some newest work exploring language models for secret-related context modeling [11, 55].

Despite achieving success in specific scenarios, these context-based approaches are often limited to open-source projects or particular secret types. Moreover, many existing tools demonstrate limited precision and recall in proprietary applications due to insufficient obfuscation handling and generalization capabilities [6, 44, 45]. This stems from their limited ability to effectively capture and fuse context and secret information. Furthermore, existing LLM-based approaches neglect intrinsic secret features and are vulnerable to semantic interference from obfuscation and cross-language scenarios, failing to demonstrate well generalization beyond datasets. A platform-agnostic tool capable of detecting diverse secret types across different environments remains needed. Therefore, we propose Secretron to enhance feature extraction and decision capabilities, along with StringGroup to systematically improve signal-to-noise ratio in context for robustness. Our experiments demonstrate consistent improvements across multiple scenarios, providing a promising direction for addressing these challenges.

## 7 Conclusion

In this work, we proposed StringGroup, a novel context extraction methodology that leverages string-based features for enhanced semantic understanding, and Secretron, a detection tool that effectively leverages context and secret information. Our approach overcomes three fundamental challenges of existing full-code learning methods: obfuscation robustness, cross-language generalization, and lengthy noisy context. Evaluation results show that Secretron significantly outperforms existing methods in complex scenarios, and in-the-wild experiments further confirm that our model captures transferable contextual semantics, generalizing to unseen secret types and languages beyond the training set. We transparently acknowledge that our current method inherently depends on string extraction heuristics and requires source-level availability. Therefore, future work will explore techniques such as binary-level analysis to alleviate these limitations. Nevertheless, the key insight that strings serve as critical semantic carriers enables our approach to overcome limitations in existing tools and may inspire future research.

## 8 Data Availability

We open our source code on our website [3]. We conducted experiments using the largest public dataset SecretBench [8] to ensure reproducibility. We are willing to share the detected in-the-wild secrets and other information with verified researchers upon reasonable request.

## Acknowledgments

This work was supported by the Natural Sciences and Engineering Research Council of Canada (NSERC) Discovery Grant no. RGCPIN-2022-03744, and by the Fonds de recherche du Québec – Nature et technologies (FRQNT) Grant no. 363482 [1]. Zhengdong Huang and Yepang Liu are also supported by Ant Group.